\documentclass[10pt,twoside]{cras}
\usepackage{amssymb,amsbsy,amsmath,amsfonts,amssymb,amscd}
\usepackage{latexsym,epsfig}
\usepackage[english,francais]{babel}
\usepackage{times}
\input utile.def
\newtheorem{e-proposition}[theorem]{Proposition}

\newtheorem{e-definition}[theorem]{Definition\rm}

\ComParit{S1296-2147}
\PIT{FLA}
\PXHY{????}
\Add{?}
\Volume{2}
\Year{2001}
\FirstPage{1}
\LastPage{??}
\AuteurCourant{Achod Aradian, \'Elie Rapha\"el and Pierre-Gilles de Gennes}
\TitreCourant{Surface flows of granular materials: some recent models} %less than 60 characters
\Journal
\Rubrique{Rubrique}{Heading}
\SousRubrique{Sous-rubrique}{Sub-Heading}
\PresentePar{First name}{NAME}
\Recu{jour mois ann\'ee}{apr\`es r\'evision}{jour mois ann\'ee}
\motscles{matériaux granulaires/ avalanches/ écoulements de 
surface}
\begin{document}
\selectlanguage{english}
\TitleOfDossier{The title of the dossier}
\TitreDeDossier{Le titre du dossier}
\title{%
Surface flows of granular materials: \\A short introduction to 
some recent models} 
\author{%
Achod Aradian~$^{\text{a}}$,\ \
\'Elie Rapha\"el~$^{\text{b}}$,\ \
Pierre-Gilles de Gennes~$^{\text{c}}$
}
\address{%
Laboratoire de Physique de la Mati\`ere Condens\'ee, Coll\`ege de 
France, URA CNRS n$^\circ$792 \\ 11, place Marcelin Berthelot, 
75231 Paris Cedex 05, France 
\begin{itemize}\labelsep=2mm\leftskip=-5mm
\item[$^{\text{a}}$]
E-mail: Achod.Aradian@college-de-france.fr
\item[$^{\text{b}}$]
E-mail: Elie.Raphael@college-de-france.fr
\item[$^{\text{c}}$]
E-mail: PGG@espci.fr
\end{itemize}
}
\maketitle
\thispagestyle{empty}
%%%%%%%%%%%%%%%%%%%%%%%%%%%%%%%%%%%%%%%%%%%%%%%%%%%%%%%%%%%%
%%%  Abstract  %%%
%%%%%%%%%%%%%%%%%%
\begin{Abstract}{%
We present a short review of recent theoretical descriptions of 
flows occuring at the surface of granular piles, and focus mainly 
on two models: the phenomenological ``BCRE'' model and the 
hydrodynamic model, based on Saint-Venant equations. Both models 
distinguish a ``static phase'' and a ``rolling'' phase inside the 
granular packing and write coupled equations for the evolutions of 
the height of each of these phases, which prove similar in both 
approaches. The BCRE description provides a very intuitive picture 
of the flow, whereas the Saint-Venant hydrodynamic description 
establishes a general and rigorous framework for granular flow 
studies.}\end{Abstract}
\enlargethispage*{2cm}
%
%
%%%%%%%%%%%%%%%%%%%%%%%%%%%%%%%%%%%%%%%%%%%%%%%%%%%%%%%%%%%%
%%%  French title  %%%
%%%%%%%%%%%%%%%%%%%%%%
\selectlanguage{french}
\begin{Ftitle}{%
Mod\`eles r\'ecents pour les écoulements surfaciques d'empilements 
granulaires }\end{Ftitle} 
%%%%%%%%%%%%%%%%%%%%%%%%%%%%%%%%%%%%%%%%%%%%%%%%%%%%%%%%%%%%
%%%  R\'esum\'e  %%%
%%%%%%%%%%%%%%%%%%%%
\begin{Resume}{%
Nous présentons une rapide revue des modèles récents décrivant les 
écoulements de surface des empilements granulaires, en nous 
concentrant plus particulièrement sur le modèle phénoménologique 
BCRE et le modèle hydrodynamique, fondé sur des équations de 
Saint-Venant. Ces deux approches font la distinction entre une 
phase statique et une phase roulante à l'intérieur du tas, et 
écrivent des équations d'évolution couplées pour l'épaisseur de 
ces deux phases qui s'avèrent similaires dans les deux cas. La 
description BCRE fournit une vision intuitive de l'écoulement, 
tandis que la description hydrodynamique établit un cadre d'étude 
rigoureux et général.}\end{Resume} 

\par\medskip\centerline{\rule{2cm}{0.2mm}}\medskip
\setcounter{section}{0}
\selectlanguage{english}
%%%%%%%%%%%%%%%%%%%%%%%%%%%%%%%%%%%%%%%%%%%%%%%%%%%%%%%%%%%%
%%%  Main text (in English)  %%%
%%%%%%%%%%%%%%%%%%%%%%%%%%%%%%%%
%
\section{Introduction}
Granular flows occur both naturally, for instance when debris roll 
down the side of a mountain or when sand avalanches on the slope 
of a desert dune, and in many industrial processes, which involve 
the conveyance of material in granular form. Describing and 
understanding this kind of flows thus appears as an issue of high 
importance from a practical point of view. Although some of the 
fundamentals laws at work inside these flows are still a matter of 
research at present (dissipation processes, existence of a 
stress-strain relation, etc.), significant progress has been made 
in the recent years with the introduction of several new 
theoretical models. This article is primarily intended to provide 
the non-specialist with a quick introduction to these recent 
models for granular flow, with a strong emphasis on two of them 
(the so-called ``BCRE'' and Saint-Venant models) which, in our 
mind, are the simplest currently available descriptions. 

We restrict ourselves to the case of dry, cohesionless granular 
media, and to situations where the granular flow is confined to a 
layer at the surface of the granular system \emph{(surface 
flows)}, i.e. situations where the typical depth of the flow is 
much smaller than its lateral dimensions (for instance, an 
avalanche at the surface of a dune). This stands in contrast with 
situations where the flow involves rearrangements deep inside the 
bulk of the system (e.g. when a heap of sand is bulldozed)---this 
latter situation is essentially the privilege of continuum 
mechanics description and requires extensive numerical 
computations. 

Roughly speaking, one can distinguish three ``generations'' of 
models describing surface flows of granular materials. The first 
generation has originated in the large amount of work devoted to 
the description of granular dynamics by researchers in the fields 
of engineering sciences and applied mechanics, and has reached a 
very convincing form in the granular flow model of Savage and 
Hutter~\cite{SavageHutter}. This very complete model writes 
classical hydrodynamic equations (incompressibility and momentum 
conservation) for the flowing material. After integration over the 
thickness of the rolling layer, depth-averaged (Saint-Venant) flow 
equations are obtained which allow the calculation of the flow 
thickness and mean velocity over the whole sample. In its 
principle, this model is restricted to the description of flows 
over \emph{fixed} bottoms of given topography (e.g. the side of a 
mountain). 

Others situations are though of interest, where the bottom is of 
the same nature and has the same properties as the flowing 
material. Consider for instance a sand avalanche on a dune: in 
this case, the position of the ``bottom'', i.e. the position of 
the border between flowing grains and material at rest, is not 
prescribed during the avalanche, since rolling grains colliding 
with initially immobile grains can either bring them into motion 
(\emph{erosion} of the bottom by the flow), or get trapped into a 
hole between two of them (\emph{deposition} of material on the 
bottom). Thus the static/mobile frontier (the ``bottom'' position) 
evolves as a dynamic variable of the problem. This distinction 
between a ``flowing phase'' and a ``static phase'', capable of 
exchanging grains through collisions processes, forms the central 
hypothesis of the second generation of models, and was developed 
by Mehta~\cite{Mehta} and by Bouchaud, Cates, Ravi Prakash and 
Edwards (also known as BCRE)~\cite{BCRECourt, BCRELong}. The next 
section is devoted to the presentation of this BCRE model. 

The third generation of models is characterized by a return to a 
hydrodynamic description (Saint-Venant equations), as in the first 
generation's Savage and Hutter model, but with the integration of 
erosion/deposition mechanisms. This approach was initiated by 
Douady \emph{et al.}~\cite{Douady}, and recently reconsidered by 
Khakhar \emph{et al.}~\cite{Khakhar} and Gray~\cite{Gray}. We 
devote section~\ref{SaintVenant} to these hydrodynamic 
descriptions. 

To this (non-exhaustive) list of models, we should also add a 
recent original proposal by Aranson and 
Tsimring~\cite{AransonCourt, AransonLong}, based on an analogy of 
granular flows with the Landau theory of phase transitions: in 
this picture, the static phase is considered as ``solid'' and the 
mobile phase as ``molten'', and an order parameter continuously 
describes the system state from molten (order parameter equal to 
zero) to solid (order parameter equal to one). The granular flow 
is then governed by a Navier-Stokes equation, but with a hybrid 
stress tensor containing both fluid-like and static-like 
constraints whose relative importance is tuned by the value of the 
order parameter. This model has been shown~\cite{AransonCourt, 
AransonLong} to reproduce experimental observations in a variety 
of situations, especially when the system is only partially 
fluidized. A fair presentation of this approach would however go 
beyond the scope of the present article, and the interested reader 
is referred to the original articles~\cite{AransonCourt, 
AransonLong}. 

As announced, the present article mainly focuses on two of the 
above-mentioned models, because of their simplicity and their 
physical content. The paper is organized as follows. In 
section~\ref{BCRE}, we present the BCRE model. In 
section~\ref{SaintVenant}, we describe the Saint-Venant 
hydrodynamic approach. In the last section (sec~\ref{Synthese}), 
we point out the relations and similarities that arise between the 
BCRE and the Saint-Venant descriptions, and conclude with a 
general discussion of our current understanding of avalanches and 
granular flows. 

\section{The BCRE model}
\label{BCRE}
The BCRE model is essentially a phenomenological description of 
granular flows, based on the hypothesis that the granular pile can 
be separated in a rolling phase (made of mobile grains) atop a 
static phase (made of immobile grains), with a sharp, well-defined 
interface. The model appeared in its original form 
in~\cite{BCRECourt, BCRELong}, and was later simplified and 
modified by Boutreux, Rapha\"el and de 
Gennes~\cite{PGGDynamiqueSuperficielle, PGGPowdersGrains, 
BoutreuxSature}. The basic quantities introduced in the BCRE 
description are depicted on Figure~\ref{FigureBCRE} (we restrict 
here to two-dimensional sandpiles): 
    \begin{figure}
    \centering
    \includegraphics*[scale=.8, clip=true, bb=2.5cm 5.5cm 18.5cm 
    11cm]{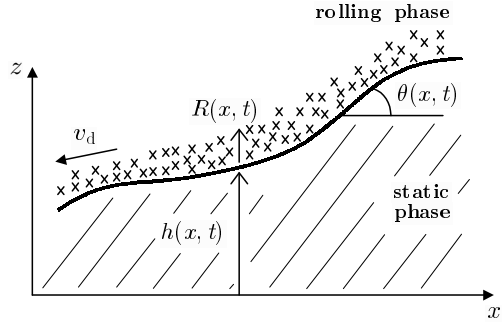} 
    \caption{Avalanche description within the BCRE model.}
    \label{FigureBCRE} 
    \end{figure} 
$h(x,t)$ is the local thickness of the static phase, at position 
$x$ and time $t$, $R(x,t)$ the local thickness of the rolling 
layer, $\theta(x,t)$ the local slope of the flowing/static 
interface ($\theta \simeq \partial h/\partial x$ for small 
slopes), and $v_{\text{d}}$ the downhill velocity of the rolling 
grains (which will be taken constant for a start). When attempting 
to describe a given situation of granular flow, one is interested 
in finding the evolution of the static and rolling height $h$ and 
$R$. In the BCRE picture, the set of equations governing the 
temporal derivatives $\partial h/\partial t$ and $\partial 
R/\partial t$ of these quantities is very 
compact~\cite{PGGDynamiqueSuperficielle, PGGPowdersGrains}: 
    \begin{subequations}
    \label{eq:SystemeBCREGeneral} 
    \begin{align}
    \label{eq:EqBCREGeneraleh}
    \frac{\partial h}{\partial t}  &= - \mathcal{E}(x,t) \,,\\
    \label{eq:EqBCREGeneraleR}
    \frac{\partial R}{\partial t}& =  
    v_{\text{d}} \frac{\partial R}{\partial x} + \mathcal{E}(x,t)\,.
    \end{align}
    \end{subequations}
In these equations, $\mathcal{E}(x,t)$ is the \emph{exchange 
term}, which represents the exchanges of grains that occur between 
static and mobile phases (dislodgement of immobile grains by 
rolling grains, or trapping of mobile grains by static grains). 
The physical meaning of the BCRE equations is very simple. 
Equation~\eqref{eq:EqBCREGeneraleh} expresses that the static 
height locally increases when material is deposited from the 
rolling phase. Equation~\eqref{eq:EqBCREGeneraleR} describes the 
evolution of the rolling thickness $R$: the first term on the 
right-hand side is a classical convection term describing the 
downhill motion of grains at velocity $v_{\text{d}}$, and the 
second term accounts for the local increase in $R$ when grains are 
entrained out of the static part. Note that in their original 
model~\cite{BCRECourt, BCRELong}, Bouchaud \emph{et al.} included 
diffusion and curvature terms which are second-order (in most 
cases, but not all), and will here be omitted. 

The next step in the BCRE description is to suggest a structure 
for the exchange term $\mathcal{E}$. Bouchaud \emph{et 
al.}~\cite{BCRECourt, BCRELong, PGGPowdersGrains} proposed the 
following expression: 
    \begin{equation}
    \label{EchangeNonSature}
    \mathcal{E}(x,t)=\gamma R (\theta - \theta_{\text{n}})\,.
    \end{equation}
This structure can be interpreted in simple physical terms as 
follows. Let $\gamma$ denote the typical collision frequency of a 
mobile grain with the static phase. Then the number of collisions 
per unit time experienced by the static phase when surmounted by a 
rolling layer of thickness $R$ is proportional to $\gamma R$. To 
evaluate the exchanges between phases, one further has to estimate 
the probability of each of these collisions of mobile grains with 
the static phase to result either in the dislodgement of a static 
grain or the trapping of the mobile grain. BCRE assume this 
probability to be mainly governed by the local slope $\theta\simeq 
\partial h/\partial x$ of the static phase. When $\theta$ equals a 
certain \emph{neutral angle} $\theta_{\text{n}}$, the probability 
of dislodgement is equal to that of trapping, and there is no net 
erosion nor deposition of grains (in other words, there is no 
amplification nor damping of the avalanche). But when $\theta > 
\theta_{\text{n}}$ (resp. $\theta 
< \theta_{\text{n}}$), there is a net erosion of grains (resp. deposition). 
For slopes that remain close to $\theta_{\text{n}}$ (usually 
around 30$^\circ$ for dry sand), one can take a linear form 
$\theta - \theta_{\text{n}}$ for this angular dependence of the 
erosion/deposition processes, and finally obtain the structure 
given in equation~\eqref{EchangeNonSature} for the global 
exchanges between static and mobile phase. 

With the above form~\eqref{EchangeNonSature} of $\mathcal{E}$, the 
general equations~\eqref{eq:SystemeBCREGeneral} then become 
    \begin{equation}
    \begin{aligned}
    \label{eq:SystemeBCRENonSature}
    \frac{\partial h}{\partial t}  &= - \gamma R (\theta - \theta_{\text{n}}) \,,\\
    \frac{\partial R}{\partial t}& =  
    v_{\text{d}} \frac{\partial R}{\partial x} + \gamma R (\theta - \theta_{\text{n}}) 
    \,.
    \end{aligned}
    \end{equation}
The evolutions of $R$ and $h$ thus appear as intimately coupled, 
through the expression of the exchange term (via its 
$R$-dependence and the local slope $\theta \simeq\partial 
h/\partial x$). Solving this set of equations (analytically or 
numerically) with appropriate boundary and initial conditions 
allows to describe the granular flow through the knowledge of $h$ 
and $R$. 

The expression~\eqref{EchangeNonSature} of the exchange term has 
however been criticized by Boutreux \emph{et al.} for flows 
thicker than a few grain diameters: in that case (assuming a a 
more or less layered flow structure), grains from the upper layers 
of the flow have no opportunity to interact and collide with the 
underlying static phase, because they are ``screened'' by lower 
flowing layers. Thus the number of collisions, and therefore the 
exchange term, cannot remain proportional to the rolling thickness 
$R$ when $R$ becomes larger than a screening length $\lambda$. 
Boutreux \emph{et al.} suggest that the exchange term must thus 
saturate for thicker flows, i.e. must be written as
    \begin{equation}
    \label{EchangeSature}
    \mathcal{E}(x,t)=v_{\text{up}} (\theta - \theta_{\text{n}}) \qquad (R > \lambda)\,,
    \end{equation}
where $v_{\text{up}}=\gamma \lambda$ has the dimensions of a 
velocity (one can indeed show that this corresponds to the 
velocity of uphill waves running at the surface of the static 
phase in avalanches, see~\cite{BoutreuxSature}). 

If we use this saturated form~\eqref{EchangeSature} of the 
exchange term into the set of governing 
equations~\eqref{eq:SystemeBCREGeneral}, we obtain the 
\emph{saturated} BCRE equations: 
    \begin{equation}
    \begin{aligned}
    \label{eq:SystemeBCRESature1}
    \frac{\partial h}{\partial t}  &= - v_{\text{up}} (\theta - \theta_{\text{n}}) \,,\\
    \frac{\partial R}{\partial t}& =  
    v_{\text{d}} \frac{\partial R}{\partial x} + v_{\text{up}} (\theta - \theta_{\text{n}}) \,.
    \end{aligned}
    \end{equation}
One pleasant feature of this saturated set of equations is that it 
is now \emph{linear} in both $h$ and $R$, and that a general 
solution can easily be formulated, in terms of uphill and downhill 
wave motions~\cite{BoutreuxSature}. 

Due to its simplicity and to the clear physical picture that it 
conveys, the BCRE model has since its creation and until recently, 
played a central role for the theoretical study and discussion of 
various situations of granular flow, either in its 
form~\eqref{eq:SystemeBCRENonSature} 
or~\eqref{eq:SystemeBCRESature1}, or in other variants for some 
complex cases. Examples of problems that have been tackled with 
the help of the BCRE description are the surface dynamics of a 
pile in a rotating drum~\cite{PGGDynamiqueSuperficielle}, the 
segregation and stratification phenomena in mixed granular 
media~\cite{BoutreuxMakse}, the formation of a sandpile from a 
point source~\cite{DorogovtsevPileFormation}, or the formation of 
ripples on a sand bed submitted to the action of 
wind~\cite{TerzidisRipples}, etc. 

However, the phenomenological nature of the BCRE description 
inevitably raises a series of important questions. (i)~The first 
of these questions is to determine whether this model respects the 
fundamental laws of mass and momentum conservation. It is easy to 
show that mass (i.e. the total number of particles $R+h$) is 
conserved, but conservation of momentum is open to doubt. (ii)~It 
is known from experiments that the velocity profile within the 
flowing layer is not constant, implying that the hypothesis 
$v_{\text{d}}=\text{const.}$ is not acceptable. How are then the 
BCRE equations to be modified? Is it enough to replace 
$v_{\text{d}}$ by the average velocity, or is there more to it? 
(iii)~A last shortcoming is due to the introduction of several 
phenomenological parameters like $\theta_{\text{n}}$ or 
$v_{\text{up}}$, which would have to be fitted to the data. One 
would naturally prefer to derive directly the expressions of these 
parameters from the fundamental quantities describing the system. 

All these questions have found a satisfactory answer with the 
recent (re-)introduction of Saint-Venant hydrodynamic descriptions 
that we are presenting in the next section.
\section{Saint-Venant hydrodynamic description}
\label{SaintVenant}
We now present the recent hydrodynamic descriptions based on 
depth-averaged (Saint-Venant) equations, which represent an 
essential step: as they start from first principle equations, they 
provide a rigourous and general framework for the study and 
understanding of surface flows of granular materials. 

The first hydrodynamic model incorporating the exchanges of grains 
between flowing and static part of the granular system is due to 
Douady \emph{et al.}~\cite{Douady}. This work was revisited very 
recently by Khakhar \emph{et al.}~\cite{Khakhar} and also by 
Gray~\cite{Gray}. For conciseness, in this presentation, we will 
not enter into all details~\cite{Douady, Khakhar, TheseAchod} and 
only give the general lines. 

The idea of these hydrodynamic models is to write conservation 
equations for mass and momentum in the flowing layer, but 
averaging all quantities across the depth of the layer 
(Saint-Venant equations). It is important to realize that, as a 
consequence, these models \emph{do not} provide any information on 
the inner structure of the flow, and can only give information on 
the flowing layer thickness and the static height.

\begin{figure}
    \centering
    \includegraphics*[scale=.8, clip=true, bb=2.5cm 3.5cm 18.5cm 
    9cm]{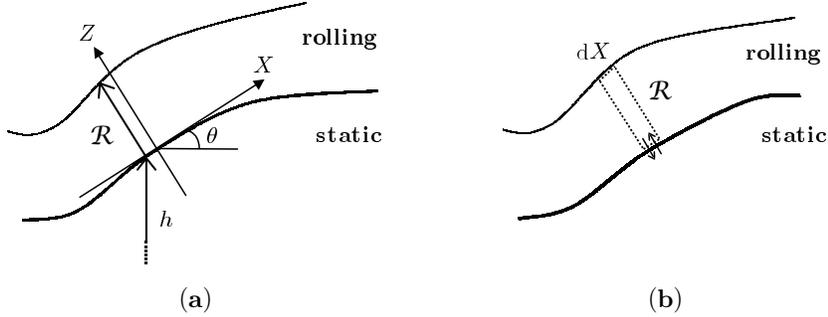} 
    \caption{(a)~Locally tangent frame $(X,Z)$. The origin $Z=0$ 
    is chosen at the rolling/static interface. (b)~Slice of fluid for
    which mass and momentum conservation are written. The arrows represent
    the exchanges between rolling and static phase.}
    \label{FigureTangent}
    \end{figure}

The BCRE equations given in the previous section were written in 
the fixed gravity frame (($x,z$) axis, see 
figure~\ref{FigureBCRE}). Here, in order to write the hydrodynamic 
conservation equations in a compact way, it will be easier to work 
temporarily in the frame ($X,Z$) locally tangent to the flow at 
each point, as shown on figure~\ref{FigureTangent}-a. The static 
height $h$ and the local slope $\theta$ are defined as before, but 
the thickness of the rolling layer is now measured in the 
direction \emph{perpendicular to the flow} (along $Z$) and will be 
noted $\mathcal{R}$ (notice the difference with $R$ which is 
reserved to the rolling thickness measured in the vertical 
direction). 
We now present the different conservation equations, and explain 
them one by one. 

The first equation is mass conservation. Following the notations 
of Khakhar \emph{et al.}~\cite{Khakhar}, we have 
    \begin{equation}
    \label{eq:ConservationMasse}
    \frac{\partial}{\partial t}(\overline\rho\, \mathcal{R})+ 
    \frac{\partial}{\partial X}(\overline{\rho v_X}\,\mathcal{R})=
    (\rho v_Z)|_{Z=0}\,,
    \end{equation}
where $\rho$ is the density of the rolling phase, and ($v_X,v_Z$) 
are the components of the velocity field in the rolling phase. The 
notation $\overline{A}$ represents the average of the quantity $A$ 
over the rolling depth: $\overline A = 
(1/\mathcal{R})\int_0^\mathcal{R} A(Z) \, \text{d} Z$. 
Equation~\eqref{eq:ConservationMasse} can be easily interpreted by 
considering a small slice of fluid of width $\text{d}X$ and height 
$\mathcal{R}$ as represented on figure~\ref{FigureTangent}-b. The 
first term on the l.h.s.\ gives the local variation of the mass of 
the slice. The second term on the l.h.s.\ gives the difference 
between input of mass at one side of the slice and output of mass 
at the other side due to the flow along the $X$-direction: this 
difference is equal to the spatial derivative of the mass flow, 
which writes $Q_X^{\text{mass}}= \int_0^\mathcal{R} \rho v_X 
\,\text{d} Z=(\overline{\rho v_X})\, \mathcal{R}$. The last term 
(on the r.h.s.) reflects that mass can also enter the slice from 
the bottom border of the slice (at $Z=0$), due to exchanges with 
the static phase: these exchanges occur along the $Z$-direction 
and the corresponding mass input is $Q_Z^{\text{mass}}=(\rho 
v_Z)|_{Z=0}$. 

The second equation~\cite{Khakhar} comes from momentum 
conservation in the $X$-direction: 
    \begin{equation}
    \label{eq:ConservationQuantiteMvmt}
    \frac{\partial}{\partial t}(\overline{\rho v_X}\,\mathcal{R})+
    \frac{\partial}{\partial X}\bigl(\overline{\rho {v_X}^2}\,\mathcal{R}\bigr)=
    \sigma_{XZ}|_{Z=0} - \overline\rho g \mathcal{R} \sin \theta\,,
    \end{equation}
where $\sigma_{XZ}$ is one of the components of the stress tensor 
{\boldmath$\sigma$} within the rolling layer and $g$ is the 
gravity field. The most general form of this conservation law 
normally includes a couple of additional terms 
(see~\cite{Khakhar}), but for simplicity and to make our point, we 
here restrict to the simplest cases where these terms are 
negligible. Here again, we can understand the meaning of each term 
of equation~\eqref{eq:ConservationQuantiteMvmt} by considering the 
slice of figure~\ref{FigureTangent}-b. The first term on the 
l.h.s.\ accounts for the local variation of momentum of the slice. 
The second term of the l.h.s.\ computes the difference between the 
momentum transferred by the flow into the slice at one side and 
out of it at the other side, as the derivative of the momentum 
flux $Q_X^{\text{moment.}} = \int_0^\mathcal{R} (\rho v_X)\cdot 
v_X\,\text{d} Z=\bigl(\overline{\rho {v_X}^2}\bigr)\, 
\mathcal{R}$. On the r.h.s.\ of the equation are gathered the 
sources of momentum due to the forces applied on the slice. The 
first term on the r.h.s.\ takes into account the friction force 
$\sigma_{XZ}|_{Z=0}$ exerted by the static phase on the bottom of 
the slice, and the second term is simply the $X$-component of the 
slice weight. 

Both equations~\eqref{eq:ConservationMasse} 
and~\eqref{eq:ConservationQuantiteMvmt} are concerned with the 
rolling thickness $R$. We now need an equation governing the 
evolution of the static height $h$. This is easily obtained by 
considering again our small slice of fluid: the mass of grains 
settling from the mobile phase to the static phase in a time 
$\text{d}t$ is $-(\rho v_Z)|_{Z=0} \cdot \text{d} t \,\text{d} X$. 
Assuming a continuous density $\rho$ across the flowing/static 
interface, this mass input induces a rise of the static height 
$\text{d}h$, so that the mass increase can also be rewritten 
$\rho|_{Z=0} \cos\theta \cdot \text{d} h \,\text{d} X$ (the 
angular factor originates in the fact that $h$ and $Z$ are at an 
angle $\theta$). Equating these two expressions of the mass, we 
obtain 
    \begin{equation}
    \label{eq:HauteurStatiqueHydro}
    \frac{\partial h}{\partial t}=- \frac{1}{\cos\theta}\, v_Z|_{Z=0} \,.
    \end{equation}

The three 
equations~\eqref{eq:ConservationMasse},~\eqref{eq:ConservationQuantiteMvmt} 
and~\eqref{eq:HauteurStatiqueHydro} form the set governing the 
granular flow in the hydrodynamic description. Let us enumerate 
the unknown quantities in these three equations: of course, 
$\mathcal{R}$ and $h$ are unknown, as is $v_Z|_{Z=0}$. But we also 
ignore the exact expressions of the density profile \emph{within} 
the flow $\rho(Z)$ and of the velocity profile $v_X(Z)$, which 
would allow the calculation of averages like $\overline{\rho v_X}$ 
or $\overline{\rho {v_X}^2}$. Thus, the set of three 
equations~\eqref{eq:ConservationMasse}--\eqref{eq:HauteurStatiqueHydro} 
is incomplete. Furthermore, there is no hope within this kind of 
depth-averaged description to gain any further knowledge on the 
internal profile of $\rho$ and $v_X$. Ideally, one would need an 
internal relation equivalent to the Navier-Stokes equation of 
ordinary fluids (and a state equation) to reach these internal 
profiles, but such a relation (if it exists!) is unknown at 
present for granular materials. 
 
To make progress, it is therefore necessary to add ``manually'' 
extra information on the physics of the system. The first usual 
step is to neglect density variations in the flow~\cite{Douady, 
Khakhar}, i.e. impose that $\rho = \overline{\rho} = 
\text{const.}$ (although this hypothesis might be a matter of 
debate if one looks for refined equations, it is useful for a 
start). 

Next, we need information on the velocity profile $v_X(Z)$. Here, 
experiments provide precious data: Rajchenbach \emph{et 
al.}~\cite{Rajchenbach1, Rajchenbach2} and Bonamy \emph{et 
al.}~\cite{BonamyVelocity} have shown that flows occurring at the 
surface of bidimensional piles (made of a monolayer of metal beads 
confined between two vertical walls) present a \emph{linear 
velocity profile}, with a vanishing velocity at the flowing/static 
interface (no slippage): 
    \begin{equation}
    \label{VelocityProfile}
    v_X(Z)=-\Gamma_0 Z\,,
    \end{equation}
where the velocity gradient $\Gamma_0$ is independent of the 
rolling thickness $\mathcal{R}$. ($\Gamma_0 \simeq \sqrt{g/d}$, 
with $d$ the grain diameter.) It is important to note, however, 
that if linear velocity profiles appear in 2D piles, the case of 
3D piles is unknown at present (to our knowledge). We note also 
that, in the different situation when the granular flow occurs at 
the surface of a fixed, inclined plane which cannot be eroded, the 
profile is non-linear with a mean velocity $\overline{v_X} \sim 
R^{3/2}$~\cite{Pouliquen, Azanza} (for recent theoretical 
proposals concerning this fact, see~\cite{Bocquet, Lemaitre}), 
whereas $\overline{v_X} \sim R$ for the linear profile of 
eq.~\eqref{VelocityProfile}. 

Finally, we need to find the expression for the friction force 
$\sigma_{XZ}|_{Z=0}$ exerted by the static phase on the rolling 
phase (eq.~\eqref{eq:ConservationQuantiteMvmt}). We choose the 
simplest form, i.e. a classical Coulomb force: 
$\sigma_{XZ}|_{Z=0}$ is taken equal to the $Z$-component of the 
weight times a (constant) dynamic friction coefficient 
$\mu_{\text{dyn}}$, and thus writes 
    \begin{equation}
    \label{FrictionForce}
    \sigma_{XZ}|_{Z=0} = \mu_{\text{dyn}} \, \rho g \mathcal{R} \cos 
    \theta\,.
    \end{equation}

We shall now incorporate all this new information into the 
hydrodynamic 
equations~\eqref{eq:ConservationMasse}--\eqref{eq:HauteurStatiqueHydro}: 
after using the assumption of constant density $\rho = 
\text{const.}$ in the flow, substituting the expressions of 
$\overline{v_X}$ and $\overline{{v_X}^2}$ as deduced 
from~\eqref{VelocityProfile}, and inserting the 
expression~\eqref{FrictionForce} of the friction force, the only 
remaining unknowns in the three equations are $h$, $\mathcal{R}$ 
and $v_Z|_{Z=0}$. One can then combine these equations together, 
and after some straightforward algebra, eventually obtain the two 
following differential equations for the static height $h$ and the 
rolling thickness $R$: 
    \begin{align}
    \label{eq:StVenantEqhRepereFixe1}
    \frac{\partial h}{\partial t}&= -  \frac{g}{\Gamma_0 \cos\theta} 
    (\sin\theta - \mu_{\text{dyn}} \cos\theta)\\
    \label{eq:StVenantEqhRepereFixe2}
    \frac{\partial R}{\partial t}&= (\Gamma_0 \cos^2\theta)\, R
    \frac{\partial R}{\partial x}
    +  \frac{g}{\Gamma_0 \cos\theta} (\sin\theta - 
    \mu_{\text{dyn}}\cos\theta)\,.
    \end{align}
Note that, in these expressions, we made the geometric 
transformation ($\mathcal{R}\to R, X \to x$) in order to come back 
from the locally tangent frame, where the hydrodynamic equations 
were initially written, to the fixed gravity frame $(x,z)$ of 
figure~\ref{FigureBCRE}, where the rolling $R$ thickness is 
measured vertically. We skip the details of this transformation 
from locally tangent to fixed frame, which can be tedious in the 
general case~\cite{Douady, Khakhar}. Here, we simply assumed, as 
happens in most practical cases, that the slope $\theta$ of the 
static phase remains roughly constant over the whole 
system~\cite{TheseAchod}: $\theta \simeq \theta_{\text{dyn}}$, 
where the angle $\theta_{\text{dyn}}$ is the ``dynamic friction 
angle'' defined by $\tan \theta_{\text{dyn}}=\mu_{\text{dyn}}$. 

Since $\theta \simeq \theta_{\text{dyn}}$, we can further simplify 
equations~\eqref{eq:StVenantEqhRepereFixe1} 
and~\eqref{eq:StVenantEqhRepereFixe2} by letting $\cos\theta 
\simeq \cos\theta_{\text{dyn}}$ and $\sin(\theta - 
\theta_{\text{dyn}}) \simeq \theta - \theta_{\text{dyn}}$. After 
some rearrangements, we finally obtain the set of equations, which 
govern the evolution of the static height $h$ and the rolling 
thickness $R$ in the Saint-Venant hydrodynamic model: 
    \begin{subequations}
    \label{eq:StVenantRepereFixeSimplifie}    
    \begin{align}
    \label{eq:StVenantRepereFixeSimplifieh}
    \frac{\partial h}{\partial t}&= -  \frac{g}{\Gamma} 
    (\theta -  \theta_{\text{dyn}}) \\
    \label{eq:StVenantRepereFixeSimplifieR}
    \frac{\partial R}{\partial t}&= \Gamma \, R
    \frac{\partial R}{\partial x} +  \frac{g}{\Gamma}(\theta - \theta_{\text{dyn}})\,, 
    \end{align}
    \end{subequations}
where we used the shorthand notation $\Gamma = \Gamma_0 \cos^2 
\theta_{\text{dyn}}$. 

In this last form, it appears that the 
equations~\eqref{eq:StVenantRepereFixeSimplifie} describing the 
flow within the Saint-Venant hydrodynamic description bear a 
striking similarity with the saturated version of the BCRE model, 
as given by equations~\eqref{eq:SystemeBCRESature1}. We will 
discuss this similarity in the next section, but we should first 
give a word of caution concerning the outcome of the Saint-Venant 
model: the final structure shown by 
equations~\eqref{eq:StVenantRepereFixeSimplifie} is directly 
dependent on the various physical assumptions that were introduced 
along the presentation (constant density, linear velocity profile, 
Coulomb friction force with a constant friction coefficient, 
near-constant static slope). However, except for special cases 
like, for instance, flows on a fixed bottom (non-linear velocity 
profile), or, possibly, very thin flowing layers (where it has 
been proposed~\cite{Douady} that the friction coefficient 
$\mu_{\text{dyn}}$ may show a dependence on $R$), we believe that 
these physical assumptions are rather robust and thus it is 
plausible that eqs.~\eqref{eq:StVenantRepereFixeSimplifie} may 
hold in a number of situations. 
\section{Relation between models and perspectives}
\label{Synthese} 
\subsection{Relation between BCRE and Saint-Venant descriptions}
After having presented the BCRE and the saint-Venant hydrodynamic 
approach in turn, we concluded that the final sets of equations 
obtained within both models (saturated 
eqs.~\eqref{eq:SystemeBCRESature1} and 
eqs.~\eqref{eq:StVenantRepereFixeSimplifie}, respectively) have 
\emph{exactly} the same structure. This outcome is especially 
interesting since these two approaches result from rather opposed 
points of view, the BCRE model being essentially phenomenological 
whereas the Saint-Venant equations originates in the application 
of hydrodynamic first principles. 

We are now in a position to answer the questions raised about the 
BCRE model at the end of section~\ref{BCRE}. First, the fact that 
a first-principle derivation leads to the same equations as the 
BCRE equations ensures that BCRE does indeed verify conservation 
of momentum. Second, the term-to-term comparison of the saturated 
equations~\eqref{eq:SystemeBCRESature1} with the saturated 
equations~\eqref{eq:StVenantRepereFixeSimplifie} allows to give 
the expressions of the phenomenological parameters $v_{\text{up}}$ 
and $\theta_{\text{n}}$ with respect to characteristic quantities 
in the problem: 
    \begin{equation}
    v_{\text{up}} =  \frac{g}{\Gamma_0 \cos^2 
    \theta_{\text{dyn}}}\,, \qquad \theta_{\text{n}} = \theta_{\text{dyn}} 
    = \arctan\mu_{\text{dyn}}\,. 
    \end{equation}
    
Finally, we see how the BCRE model would have to be modified to 
take into account that the actual velocity profile in the flow is 
linear (instead of the model's original assumption of a constant 
downhill velocity $v_{\text{d}}$): one simply has to take the 
following expression 
 $v_{\text{d}} = v_{\text{d}}(R)=\Gamma_0 \cos^2 \theta_{\text{dyn}} R$, 
which is equal (within angular factors) to the depth-averaged 
velocity of the flow. However, though this process seems rather 
intuitive, one should not infer that the BCRE model can be 
extended to \emph{any} type of velocity profile by simply 
introducing the mean velocity in place of the downhill velocity 
$v_{\text{d}}$: this deceptively simple conclusion proves 
erroneous in the general case, as first pointed out by Douady 
\emph{et al.}~\cite{Douady}. In fact, if the velocity profile in 
the flowing layer is not linear (nor constant), the hydrodynamic 
approach proves that the 
$h$-equation~\eqref{eq:StVenantRepereFixeSimplifieh} includes 
several additional terms and becomes significantly more complex. 
(And the simple result for linear, or constant, velocity can 
actually easily be seen to come as the result of a favorable 
cancellation of terms\ldots). 

One of the great merits of the BCRE model is to give a very 
intuitive picture of the granular flow through the idea of grain 
``exchanges'' between the static and rolling phases occurring by 
dislodgement and trapping mechanisms, and it is indeed very 
satisfiying to see this picture confirmed by the more rigourous 
hydrodynamic approach. In our mind, the BCRE description remains 
therefore useful as a way of thinking and ``visualizing'' granular 
surface flows. On the other hand, the hydrodynamic description is 
both more general and more rigourous, and thus constitutes, as of 
today, a reference tool for the study of granular flows. 
\subsection{Perspectives on the surface flows of granular materials}
We hope to have shown in the present short review that the 
theoretical description of surface flows of granular materials has 
made significant advances in the last fifteen years, and that we 
presently are in possession of reasonably efficient and reliable 
models. 

\begin{figure}
    \centering
    \includegraphics*[scale=.9, clip=true, bb=4cm 4.5cm 17cm 
    9cm]{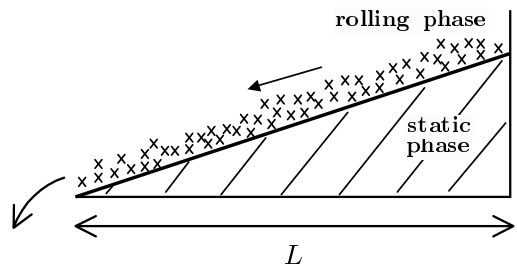} 
    \caption{An ``open'' pile of grains with an avalanche taking place at its surface.}
    \label{OpenPile}
    \end{figure}

It is worth mentioning that recent experimental results by Khakhar 
\emph{et al.}~\cite{Khakhar} have confirmed some theoretical 
predictions made with the help of the governing 
equations~\eqref{eq:StVenantRepereFixeSimplifie}: in 
ref.~\cite{Aradian}, it had been predicted that when an avalanche 
occurs at the surface of an ``open'' pile of grains (i.e. a pile 
where the grains are free to fall at the bottom end), as shown on 
figure~\ref{OpenPile}, 
part of the rolling profile $R(x,t)$ has a parabolic shape (and 
therefore, the maximum thickness of the avalanche scales as the 
square root of the system size, i.e.\ $R_{\text{max}} \sim 
\sqrt{L}$). This is the same type of parabolic profile that has 
indeed been observed experimentally by Khakhar \emph{et 
al.}~\cite{Khakhar}. 

Yet, at present, our understanding of the surface flows of 
granular flows remains fragmented and many elements lack to form a 
coherent and global picture. We would like to conclude this 
presentation by pointing out two of the least understood points. 

Perhaps one of the most important issue is now to understand how 
an avalanche \emph{starts}. The initiation process still remains 
obscure, and one can think of several mechanisms. For instance, 
the avalanche may start by a delocalized mechanism: above a 
certain maximum stability angle, the top layer of the pile is 
destabilized, starts to slide as a whole and is rapidly fluidized 
by collisions, hence forming a thin initial layer of rolling 
grains over the surface. On the contrary, one may rather favor a 
localized mechanism: at some moment, the most unstable grain(s) 
start to roll, and progressively disturb their neighborhood, 
thereby creating avalanching regions which spread around. Recent 
experiments by Rajchenbach~\cite{RajchenbachTriggering} seem to 
support this second scenario. We can even think that the 
delocalized and localized pictures are not completely 
incompatible: the delocalized scenario might be considered as a 
``coarse-grained'' view of the avalanche initiation process, not 
valid at the grain scale where localized nucleation mechanisms 
predominate, but sensible after a short time for which the whole 
surface has finally been disturbed and put into motion. It would 
be interesting to have an estimate of this ``complete 
destabilization'' time of the surface originating from a few, 
localized, triggering points. We may also note that, in describing 
such early processes of initiation, the granular flow model by 
Aranson and Tsimring~\cite{AransonCourt, AransonLong} (mentioned 
in the Introduction) might be better suited, as its very principle 
is to allow for \emph{partially} fluidized states. 

Another significant (and very difficult) issue is to get a better 
understanding of the internal rheology of these surface flows: 
beyond the depth-averaged descriptions presented here, one would 
for instance like to better understand why linear velocity 
profiles emerge in flows taking place at the surface of piles, and 
why non-linear profiles arise in flows over fixed bottoms. We may 
mention here two arguments that have been proposed in the 
literature: the first is due to Komatsu \emph{et 
al.}~\cite{Komatsu} who noticed that when a flow takes place at 
the surface of a pile, the static phase in fact undergoes a slow 
creeping motion; these authors suggest that the supression of this 
creeping motion when the flow occurs on a \emph{rigid} bottom may 
be responsible for the change in the nature of the velocity 
profile between these two types of experiments. The second 
proposal, by Bonamy \emph{et al.}~\cite{BonamyClusters} is very 
recent: these authors have observed experimentally that the 
texture of granular flows is strongly inhomogeneous, with 
``solid'' clusters of grains embedded within the flowing layer. 
For flows at the surface of a pile, it was found that the 
size-distribution of these clusters follows a power-law with sizes 
ranging from the grain size to the flowing layer thickness. Bonamy 
\emph{et al.}~\cite{BonamyClusters} then suggest that flows over 
fixed bottoms should display a very different cluster size 
distribution, which may be at the origin of the very different 
rheological behaviour observed in these systems as compared to 
piles. 

Much remains to be done before we reach a comprehensive theory of 
surface flows of granular materials. However, the rapid pace 
sustained within this field on both the experimental and 
theoretical sides, over the last few years, may be interpreted as 
a good sign in the exploration of one of the (many) intriguing 
aspects of granular matter. 
%
%
%
%
%%%%%%%%%%%%%%%%%%%%%%%%%%%%%%%%%%%%%%%%%%%%%%%%%%%%%%%%%%%%
%%%  Acknowledgements  %%%
%%%%%%%%%%%%%%%%%%%%%%%%%%
\Acknowledgements{The authors would like to thank J.~Duran for his 
constant interest in their work.} 
%%%%%%%%%%%%%%%%%%%%%%%%%%%%%%%%%%%%%%%%%%%%%%%%%%%%%%%%%%%%
%%%  Bibliography  %%%
%%%%%%%%%%%%%%%%%%%%%%%

%
\end{document}